\documentclass[10pt]{llncs} \usepackage{amsmath} \usepackage{amstext}
\usepackage{amsbsy}

\usepackage{makeidx} 
\usepackage{latexsym,amsmath,amstext,amsfonts,amsbsy,amssymb,graphicx,hyperref}
\usepackage{algorithm,algorithmic,prettyref}

\newcommand{\code}[1]{\begin{small}\texttt{#1}\end{small}}

\title{Diverse Consequences of Algorithmic Probability}

\titlerunning{Consequences of ALP} \author{Eray \"Ozkural}
\institute{Computer Engineering Department, Bilkent University,
  Ankara, Turkey} 
\date{\today}

\begin{document}

\maketitle
\begin{abstract}
  We reminisce and discuss applications of algorithmic probability to
  a wide range of problems in artificial intelligence, philosophy and
  technological society. We propose that Solomonoff has effectively
  axiomatized the field of artificial intelligence, therefore
  establishing it as a rigorous scientific discipline. We also relate
  to our own work in incremental machine learning and philosophy of
  complexity.
\end{abstract}

\section{Introduction}

Ray Solomonoff was a pioneer in mathematical Artificial Intelligence
(AI), whose proposal of Algorithmic Probability (ALP) has led to
diverse theoretical consequences and applications, most notably in AI.
In this paper, we try to give a sense of the
significance of his theoretical contributions, reviewing the essence
of his proposal in an accessible way, and recounting a few, seemingly
unrelated, diverse consequences which, in our opinion, hint towards a
philosophically clear world-view that has rarely been acknowledged 
by the greater scientific community. That is to say, we try to give
the reader a glimpse of what it is like to consider the consequences
of ALP, and what ideas might lie behind the theoretical model,
 as we imagine them.

Let $M$ be a reference machine which corresponds to a 
universal computer\footnote{Optionally, it can be probabilistic to
  deal with general induction problems, i.e., it has access to a 
random number generator \cite[Section 4]{solomonoff-progress}.} 
with a prefix-free code.
In a prefix-free code, no code is a prefix of
another. This is also called a self-delimiting code, as most
reasonable computer programming languages are.
Solomonoff inquired the probability that an output string $x$ is
generated by $M$ considering the whole space of possible programs.  By
giving each program bitstring $p$ an a priori probability of
$2^{-|p|}$, we can ensure that the space of programs meets the
probability axioms (by the extended Kraft inequality \cite{chaitin1975}).  In other
words, we imagine that we toss a fair coin to generate each bit of a
random program. This probability model of programs entails the
following probability mass function (p.m.f.) for strings $x \in
\{0,1\}^*$:
\begin{equation}
\label{eq:alp}
  P_M(x) = \sum_{M(p) = x*} 2^{-|p|}
\end{equation}
which is the probability that a random program will output a prefix of
$x$. $P_M(x)$ is 
called the \emph{algorithmic probability} of $x$ for it assumes
the \emph{definition} of program based probability. We use $P$ when
$M$ is clear from the context to avoid clutter.

\section{Solomonoff Induction}

Using this probability model of bitstrings, one can make
predictions. Intuitively, we can state that it is impossible to
imagine intelligence in the absence of any prediction ability: purely
random behavior is decisively non-intelligent. Since, $P$ is a
universal probability model, it can be used as the basis of universal
prediction, and thus intelligence.  Perhaps, Solomonoff's most
significant contributions were in the field of AI, as he envisioned a
machine that can learn anything from scratch. Reviewing his early
papers such as \cite{sol56,sol57}, we see that he has established the
theoretical justification for machine learning and data mining
fields. Few researchers could ably make claims about universal 
intelligence as he did. 
Unfortunately, not all of his ideas have reached fruition in practice;
yet there is little doubt that his approach was the correct basis for
a \emph{science} of intelligence.

His main proposal for machine learning is inductive inference
\cite{alp1,alp2} circa 1964, for a variety of problems such as sequence
prediction, set induction, operator induction and grammar induction
\cite{solomonoff-threekinds}. Without much loss of generality, we can
discuss sequence prediction on bitstrings. Assume that there is a
\emph{computable} p.m.f. of bitstrings $P_1$.
Given a bitstring $x$ drawn from $P_1$, we can define the conditional
probability of the next bit simply by normalizing
\prettyref{eq:alp} \cite{solomonoff-threekinds}. 
Algorithmically, we would have to approximate \prettyref{eq:alp} by
finding short programs that generate $x$ (the shortest of which is the
most probable). In more general induction, we run
all models in parallel, quantifying fit-to-data, weighed by the 
algorithmic probability of the model, to find the best models and
construct distributions \cite{solomonoff-threekinds}; the
common point being determining good models with high a priori probability. 
Finding the shortest program in general is
\emph{undecidable}, however, Levin search \cite{levin-search} can be
used for this purpose. There are two important results about
Solomonoff induction that we shall mention here. First, Solomonoff
induction converges very rapidly to the real probability
distribution. The convergence theorem shows that the expected total
square error is related only to the algorithmic complexity of $P_1$,
which is independent from $x$. The following bound
\cite{solomonoff-theoryandapps} is discussed at length in
\cite{solcomplexity} with a concise proof:
\begin{equation}
\label{eq:convergence}
  E_P \left[ \sum_{m=1}^n{(P({a_{m+1}=1}|a_1a_2...a_m) -
      P_1({a_{m+1}=1}|a_1a_2...a_m))^2}  \right] 
  \leq - \frac{1}{2} \ln{P(P_1))}
\end{equation}
This bound characterizes the divergence of the ALP solution from the
real probability distribution $P_1$.  $P(P_1)$ is the a priori
probability of $P_1$ p.m.f.  according to our universal distribution
$P_M$. On the right hand side of \prettyref{eq:convergence},
 $-\ln{P_M(P_1)}$ is roughly $k\ln{2}$ where $k$ is the
Kolmogorov complexity of $P_1$ (the length of the shortest program
that defines it), thus the total expected error is bounded by a constant,
which guarantees that the error decreases very rapidly as example size
increases.  Secondly, there is an optimal search algorithm to
approximate Solomonoff induction, which adopts Levin's universal
search method to solve the problem of universal induction
\cite{levin-search,solomonoff-optimum}.  Universal search procedure
time-shares all candidate programs according to their a priori
probability with a clever watchdog policy
to avoid the practical impact of the undecidability of the halting
problem \cite{solomonoff-optimum}.
The search procedure starts with a time limit $t=t_0$, in its iteration tries
all candidate programs $c$ with a time limit of $t.P(c)$, and while a
solution is not found, it doubles the time limit $t$. 
The time $t(s)/P(s)$ for a solution program $s$
taking time $t(s)$ is called the Conceptual Jump Size (CJS), and it
is easily shown that Levin Search terminates in at most $2.\text{CJS}$
time. To obtain alternative solutions, one may keep running after the
first solution is found, as there may be more probable solutions that
need more time.  The optimal solution is computable only in the limit, 
which turns out to be a desirable property of Solomonoff
induction, as it is \emph{complete} and uncomputable \cite[Section
2]{alpstrong}. An explanation of Levin's universal search
procedure and its application to Solomonoff induction may be found in
\cite{levin-search,solomonoff-optimum,solomonoff-incremental}.

\vspace*{-4pt}
\section{The Axiomatization of Artificial Intelligence}
\vspace*{-2pt}

We believe in fact that Solomonoff's work was seminal in that he has
single-handedly \emph{axiomatized} AI, discovering the minimal
necessary conditions for any machine to attain general intelligence
(based on our interpretation of \cite{solomonoff-progress}).

Informally, these axioms are:
\begin{description}
\item[AI0] AI must have in its possession a universal computer M
  (Universality).
\item[AI1] AI must be able to learn any solution expressed in M's code
  (Learning recursive solutions).
\item[AI2] AI must use probabilistic prediction (Bayes' theorem).
\item[AI3] AI must embody in its learning a principle of induction
  (Occam's razor).
\end{description}

While it may be possible to give a more compact characterization,
these are ultimately what is necessary for the kind of general
learning that Solomonoff induction achieves. ALP can be seen as a
\emph{complete} formalization of Occam's razor (as well as Epicurus's principle)
\cite{rathmaner-philosophical} and thus serve as the foundation of
universal induction, capable of solving all AI problems of
significance. The axioms are important because they allow us to assess
whether a system is capable of general intelligence or not.

Obviously, AI1 entails AI0, therefore AI0 is redundant, and can be
omitted entirely, however we stated it separately only for historical
reasons, as one of the landmarks of early AI research, in retrospect,
was the invention of the universal computer, which goes back to
Leibniz's idea of a universal language (characteristica universalis)
that can express every statement in science and mathematics, and has
found its perfect embodiment in Turing's research
\cite{davis-universalcomputer,turing-computable}. A related
achievement of early AI was the development of LISP,
a universal computer based on lambda calculus (which is a functional model of
computation) that has shaped much of early AI research.

See also a recent survey about inductive inference \cite{dowe-survey}  
with a focus on Minimum Message Length (MML) principle introduced in
1968 
\cite{WallaceBoulton:1968}. MML principle is also a formalization of
induction developed within the framework of classical information
theory, which establishes a trade-off between model complexity and
fit-to-data by finding the minimal message that encodes both the model and the data~\cite{Wallace:05}. This trade-off is quite similar to the earlier
forms of induction that Solomonoff developed, however independently
discovered.
Dowe points out that Occam's razor means
choosing the simplest single theory when data is equally matched,
which MML formalizes perfectly (and is functional otherwise in the case of inequal
fits) while Solomonoff induction maintains a
mixture of alternative solutions \cite[Sections 2.4 \& 4]{dowe-survey}.
On the other hand, the diversity of solutions in ALP is seen as
desirable by Solomonoff himself \cite{alpstrong}, and in a recent
philosophical paper which illustrates how Solomonoff induction
dissolves various philosophical objections to induction 
\cite{rathmaner-philosophical}. 
Nevertheless, it is well
worth mentioning that Solomonoff induction (formal theory published in 1964 \cite{alp1,alp2}),
 MML (1968), and Minimum
Description Length \cite{barron2000} formalizations, as well as Statistical Learning
Theory \cite{vapnik-statistical} (initially developed in 1960), all
provide a principle of induction (AI3). However, it was
Solomonoff who first observed the importance of universality for AI
(AI0-AI1). The plurality of probabilistic approaches to induction
supports the importance of AI3 (as well as hinting that diversity of
solutions may be useful). AI2, however, does not require much explanation. 
Some objections to Bayesianism are answered using MML in \cite{Dowe2007}. 
Please also see an intruging paper by Wallace and Dowe
\cite{wallace99} on the relation between MML and
Kolmogorov complexity, which states that Solomonoff induction is
tailored to prediction rather than inference, and recommends  
non-universal models in practical work, therefore becomes incompatible with  
the AI axioms (AI0-AI1). Ultimately, empirical work will illuminate 
whether our AI axioms should be adopted, or  more restrictive
models are sufficient for universal intelligence; therefore such
alternative viewpoints must be considered. In addition to this, 
Dowe discusses the relation between inductive inference and
intelligence, and the requirements of intelligence as we do 
elsewhere \cite[Section 7.3]{dowe-survey}.  Also relevant is an
adaptive universal intelligence test that aims to measure the
intelligence of any AI agent, and discusses various definitions
of intelligence \cite{orallo2010}.

\vspace*{-4pt}
\section{Incremental Machine Learning}

In solving a problem of induction, the aforementioned search methods
suffer from the huge computational complexity of trying to compress
the entire input. For instance, if the complexity of the p.m.f.  $P_1$
is about $400$ bits, Levin search would take on the order of $2^{400}$
times the running time of the solution program, which is infeasible
(quite impossible in the observed universe).  Therefore, Solomonoff has
suggested using an \emph{incremental} machine learning algorithm,
which can re-use information found in previous solutions
\cite{solomonoff-incremental}.

The following argument illustrates the situation more clearly.  Let
$P_1$ and $P_2$ be the p.m.f.'s corresponding to a training sequence
of two induction problems (any of them, not necessarily sequence
prediction, to which others can be reduced easily) with data $<d_1,
d_2>$.  Assume that the first problem has been solved (correctly) with
universal search. It has taken at most $2.\text{CJS}_1 =
2.t(s_1)/P(s_1)$ time. If the second problem is solved in an
\emph{incremental} fashion, making use of the information from $P_1$,
then the running time of discovering a solution $s_2$ for $d_2$
reduces, depending on the success of \emph{information transfer}
across problems. Here, we quantify how much in familiar probabilistic
terms.

In \cite{solcomplexity}, Solomonoff describes an information theoretic
interpretation of ALP, which suggests the following entropy function:
\begin{equation}
  H^*(x) = -\log_2{P(x)}
\end{equation}
This entropy function has perfect sub-additivity of information
according to the corresponding conditional entropy definition:
\begin{align}
  P(y|x) &= \frac{P(x,y)}{P(x)}\\
  H^*(y|x) &= -\log_2{P(y|x)}\\
  H^*(x,y) &= H^*(x) + H^*(y|x)
\end{align}
This definition of entropy thus does not suffer from the additive
constant terms as in Chaitin's version. We can instantly define mutual
entropy:
\begin{equation}
  H^*(x:y) = H^*(x) + H^*(y) - H^*(x,y) = H^*(y) - H^*(y|x)
\end{equation}
which trivially follows.

A KUSP machine is a universal computer that can store data and methods
in additional storage.  In 1984, Solomonoff observed that KUSP
machines are especially suitable for incremental learning
\cite{solomonoff-optimum}. In our work \cite{DBLP:conf/agi/Ozkural11}
we found that, the incremental learning approach was indeed useful (as
in the preceding OOPS algorithm\cite{oops}). Here is how we
interpreted incremental learning. After each induction problem, the
p.m.f. $P$ is updated, thus for every new problem a new probability
distribution is obtained. Although we are using the same $M$ reference
machine for trial programs, we are referring to \emph{implicit} KUSP
machines which store information about the experience of the machine
so far, in subsequent problems. In our example of two induction
problems, let the updated $P$ be called $P'$, naturally there will be
an update procedure which takes time $t_u(P, s_1)$. Just how much time
can we expect to save if we use incremental learning instead of
independent learning? First, let us write the time bound $2.t(s)/P(s)$
as $t(s).2^{H^*(s)+1}$. If $s_1$ and $s_2$ are not algorithmically
independent, then $H^*(s_2|s_1)$ is smaller than $H^*(s_2)$.
Independently, we would have $t(s_1).2^{H^*(s_1)+1} +
t(s_2).2^{H^*(s_2)+1}$, together, we will have, in the best case
$t(s_1).2^{H^*(s_1)+1} + t(s_2).2^{H^*(s_2|s_1)+1}$ for the search
time, assuming that recalling $s_1$ takes no time for the latter
search task (which is an unrealistic assumption). Therefore in total,
the latter search task can accelerate $2^{H^*{(s1:s2)}}$ times, and we
can save $t(s_2).2^{H^*(s_2)+1}(1 - 2^{-H^*{(s1:s2)}}) -t_u(P,s_1)$
total time in the best
case (only an upper bound since we did not account for recall
time). Note that the maximum temporal gain is related to both how much
mutual information is discovered across solutions (thus $P_i$'s), and how much time the update procedure takes. Clearly, if
the update time dominates overall, incremental learning is in vain.
However, if updates are effective and efficient, there is enormous
potential in incremental machine learning.

During the experimental tests of our Stochastic Context Free Grammar
based search and update algorithms \cite{DBLP:conf/agi/Ozkural11}, we
have observed that in practice we can realize fast updates, and we can
still achieve actual code re-use and tremendous speed-up.  Using only
$0.5$ teraflop/sec of computing speed and a reference machine choice
of R5RS Scheme~\cite{r5rs}, 
we solved $6$ simple deterministic operator induction
problems in $245.1$ seconds.  This running time is compared to $7150$
seconds without any updates.  Scaled to human-level processing
speed of $100$ teraflop/sec, our system would learn and solve the
entire training sequence in $1.25$ seconds, which is (arguably) better
than most human students. In one particular operator induction problem
(fourth power, $x^4$), we saw actual code re-use: \code{(define (pow4
  x ) (define (sqr x ) (* x x)) (sqr (sqr x ) ))}, and an actual
speedup of $272$. The gains that we saw confirmed the incremental
learning proposals of Solomonoff, mentioned in a good number of his
publications, but most clearly in
\cite{solomonoff-optimum,solomonoff-incremental,solomonoff-progress}.
Based on our work and the huge speedup observed in OOPS for a shorter
training sequence~\cite{oops}, we have come to believe that
incremental learning has the epistemological status of an additional
AI axiom:
\begin{description}
\item[AI4] AI must be able to use its previous experience to speed up
  subsequent prediction tasks (Transfer Learning).
\end{description}
This axiom is justified by observing that many universal induction
problems are completely unsolvable by a system that does not have the
adequate sort of \emph{algorithmic} memory, regardless of the search
method.

The results above may be contrasted with inductive programming
approaches, since we predicted deterministic functions. 
One of the earliest and most successful inductive programming systems
is ADATE, which is optimized for a more specific purpose.  
ADATE  system has yielded impressive results in an ML variant by user
supplied primitives and constraining candidate programs \cite{adate}.
Universal representations have been investigated in inductive logic programming
as well \cite{muggleton1994learnability}, however U-learning
unfortunately lacks the extremely accurate generalization of
Solomonoff  induction.
 It has been shown that incremental learning
is useful in the inductive programming framework
\cite{ferri2001incremental}, which supports our observation of the
necessity of incremental machine learning. Another relevant
work is a typed
higher-order logic knowledge representation scheme based on term
representation of individuals and a rich representation language
encompassing many abstract data types \cite{bowers2001knowledge}.
A recent survey on inductive programming may be found in
\cite{kitzelmann2010inductive}.

We should also account our brief correspondence with Solomonoff. We
expressed that the prediction algorithms were powerful but it seemed
that memory was not used sufficiently.  Solomonoff responded by
mentioning the potential stochastic grammar and genetic programming
approaches that he was working on at the time.  Our present research
was motivated by a problem he posed during the discussions of his
seminars in Turing Days '06 at Bilgi University, Istanbul: 
``We can use grammar induction for updating a
stochastic context free grammar, but there is a problem. We already
know the grammar of the reference machine.''.  We designed our
incremental learning algorithms to address this particular
problem\footnote{We 
occassionally corresponded via e-mail. Before the
  AGI-10 conference, he had reviewed a draft of my paper, and he had
  commented that the ``learning programming idioms'' and 
  ``frequent subprogram mining''
  algorithms were interesting, which was all the encouragement I
  needed. The last e-mail I received from him was
  on 11/Oct/2009. I regretfully learnt that he passed away a month 
  later. His independent character and true scientific spirit
  will always be a shining beacon for me.}.
Solomonoff has also guided our research by making a valuable
suggestion, that it is more important to show whether incremental
learning works over a sequence of simpler problems than solving a 
difficult problem. We have in addition investigated the use of PPM
family of compressors  following his proposal,
but as we expected, they were not sufficient for guiding LISP-like
programs, and would require too many changes. Therefore, we proceeded
directly to the simplest kind of guiding p.m.f. that would work for
Scheme, as we preferred not to work on assembly-like languages for
which PPM might be appropriate, since, in our opinion, high-level
languages embody more technological progress (see also
~\cite{Looks:2007} which employs a Scheme subset). 
Colorfully speaking, inventing a
functional form in assembly might be like re-inventing the wheel. 
However, in general, it would not be trivial for the induction system to invent
syntax forms that compare favorably to LISP, especially during
preliminary training. Therefore, much intelligence is already present
in a high-level universal computer (AI0) which we simply take
advantage of.

\vspace*{-4pt}
\section{Cognitive Architecture}

\vspace*{-2pt} 
Another important discussion is whether a cognitive architecture is
necessary.  The axiomatic approach was seen counter-productive by some
leading researchers in the past. However, we think that their opinion
can be expressed as follows: the minimal program that realizes these
axioms is not automatically intelligent, because in practice an
intelligent system requires a good deal of algorithmic information to
take off the ground. This is not a bad argument, since obviously, the
human brain is well equipped genetically. However, we cannot either
rule out that a somewhat compact system may achieve human-level
general intelligence.  The question therefore, is whether a simply
described system like $\text{AIXI}$~\cite{Hutter:07aixigentle} 
(an extension of Solomonoff induction to reinforcement learning)
is sufficient \emph{in practice}, or there is a need for a
modular/extensible cognitive architecture that has been designed in
particular ways to promote certain kinds of mental growth and
operation. Some proponents of general purpose AI research think that
such a cognitive architecture is necessary, e.g.,  OpenCog
\cite{Goertzel09a}. Schmidhuber has suggested the famous G\"{o}del
Machine which has a mechanical model of machine consciousness
\cite{Schmidhuber:09gm}.  Solomonoff himself has proposed early on in
2002, the design of Alpha, a generic AI architecture which can
ultimately solve free-form time-limited optimization problems
\cite{solomonoff-incremental}. Although in his later works, Solomonoff
has not made much mention of Alpha and has instead focused on the
particulars of the required basic induction and learning capability,
nonetheless his proposal remains as one of the most extensible and
elegant self-improving AI designs. Therefore, this point is open to
debate, though some researchers may want to assume another, entirely
optional, axiom:
\begin{description}
\item[AI5] AI must be arranged such that self-improvement is feasible
  in a realistic mode of operation (Cognitive Architecture).
\end{description}
It is doubtful for instance whether a combination of incremental
learning and $\text{AIXI}$ will result in a
practical reinforcement learning agent. Neither is it well understood
whether autonomous systems with built-in utility/goal functions are
suitable for all practical purposes. We anticipate that such questions
will be settled by experimenters, as the complexity of interesting
experiments will quickly overtake theoretical analysis. 

We do not consider 
human-like behavior, or a robotic body, or an autonomous AI design, such
as a goal-driven or reinforcement-learning agent, essential to
intelligence, hence we did not propose autonomy or embodiment as an axiom. 
Solomonoff has commented likewise on the preferred target applications    
\cite{solomonoff-pastandfuture}:
\begin{quote}
To start, I'd like to define the scope of my interest in A.I. I am not
particularly interested in simulating human behavior. I am interested
in creating a machine that can work very difficult problems much
better and/or faster than humans can -- and this machine should be
embodied in a technology to which Moore's Law applies. I would like it
to give a better understanding of the relation of quantum mechanics to
general relativity. I would like it to discover cures for cancer and
AIDS. I would like it to find some very good high temperature
superconductors. I would not be disappointed if it were unable to pass
itself off as a rock star.
\end{quote}

\vspace*{-4pt}
\section{Philosophical Foundation and Consequences}

\vspace*{-2pt}
Solomonoff's AI theory is founded on a wealth of philosophy.
 Here, we shall briefly revisit the philosophical foundation of ALP
 and point out some of its 
philosophical consequences. In his posthumous publication,
Solomonoff mentions the inspiration for some of his work:
Carnap's idea that the state of the world
can be represented by a finite bitstring (and that science predicts future
bits with inductive inference),
 Turing's universal computer (AI0) as
communicated by Minsky and McCarthy, and Chomsky's generative grammars
\cite{alpstrong}. The discovery of ALP is described by Solomonoff
in quite a bit of detail in \cite{Sol:97discovery}, which relates his
discovery to the background of many prominent thinkers and
contributors. 
Carnap's empiricism seems to have been a highly influential factor in
Solomonoff's research as he sought to find how science is carried out, rather than
particular scientific findings; and ALP is a satisfactory solution to
Carnap's program of inductive inference \cite{rathmaner-philosophical}.

Let us then recall some philosophically relevant aspects of ALP
discussed in the most recent publications of Solomonoff. First, the
exact same method is used to solve both mathematical and scientific
problems. This means that there is no fundamental epistemological
difference between these problems; our interpretation is that, this is
well founded only when we observe that mathematical problems
themselves are computational or linguistic problems, in practice
mathematical problems can be reduced to particular computational
problems, and here is why the same method works for both kinds of
problems. Mathematical facts do not preside over or precede physical
facts, they themselves are solutions of physical problems ultimately
(e.g., does this particular kind of machine halt or not?).  And the
substance of mathematics, the lucid sort of mathematical language and
concepts that we have invented, can be fully explained by Solomonoff
induction, as those are the kinds of \emph{useful programs}, which
have aided an intellect in its training, and therefore are retained as
\emph{linguistic} and \emph{algorithmic} information. The subjectivity
and diversity aspects of ALP \cite[Sections 3 \& 4]{alpstrong}
fully explain why there can be multiple and almost equally productive
foundations of mathematics, as those merely point out somewhat equally
useful formalisms invented by different mathematicians. There is
absolutely nothing special about ZFC theory, it is just a formal
theory to explain some useful procedures that we perform in our heads,
i.e., it is more like the logical explanation of a set module in a
functional programming language than anything else, however, the
operations in a mathematician's brain are not visible to their owner,
thereby leading to useless Platonist fantasies of some mathematicians
owing to a dearth of philosophical imagination.  Therefore, it does
not matter much whether one prefers this or that formalization of set
theory, or category theory as a foundation, unless that choice
restricts success in the solution of future scientific
problems. Since, such a problematic scientific situation does not seem
to have emerged yet (forcing us to choose among particular
formalizations), the diversity principle of ALP forces us to retain
them all. That is to say, subscribing to the ALP viewpoint has the
unexpected consequence that we abandon both Platonism and
Formalism. There is a meaning in formal language, in the manner which
improves future predictions, however, there is not a single a priori
fact, in addition to empirical observations, and no
such fact is ever needed to conduct empirical work, except a proper
realization of axioms A1--A3 (and surely no sane scientist would accept
that there is a unique and empty set that exists in a hidden order of
reality).  When we consider these axioms, we need to understand
the universality of computation, and the principled manner in which
we have to employ it for reliable induction in our scientific inquiries. The only
physically relevant assumption is that of the computability of the
distributions which generate our empirical problems (regardless of whether the
problem is mathematical or scientific), and the choice of a universal
computer which introduces a \emph{necessary} subjectivity. The
computability aspect may be interpreted as \emph{information
  finitism}, all the problems that we can work with should have finite
entropy. Yet, this restriction on disorder is not at all limiting, for
it is hardly conceivable how one may wish to solve a problem of
actually infinite complexity. Therefore, this is not much of an
assumption for scientific inquiry, especially given that both quantum
mechanics and general relativity can be described in computable
mathematics (see for instance \cite{bridges-constructivequantum} about
the applicability of computable mathematics to quantum mechanics).
And neither can one hope to find an example of a single scientifically
valid problem in any textbook of science that requires the existence
of distributions with infinite complexity to solve.

With regards to general epistemology, ALP/AIT may be seen as largely 
incompatible with non-reductionism. Non-reductionism is quite
misleading in the manner it is usually conveyed. Instead, we
must seek to understand irreducibility in the sense of AIT, of
quantifying algorithmic information, which allows us to reconcile the
concept of irreducibility with physicalism (which we think every empiricist 
should accept) \cite{ozkural-compromise}.  In particular, we can
partially formalize the notion of knowledge by mutual information
between the world and a brain. Our paper
 proposed a physical solution to the problem of
determining the most ``objective'' universal computer: it is the
universe itself. If digital physics were true, this might be for
instance a particular kind of graph automata, or if quantum mechanics
were the basis, then a universal quantum computer could be used; 
however, for many tasks using such a low-level computer might be
extraordinarily difficult. We also argued that extreme
non-reductionism 
leads to arguments from ignorance such as ontological dualism, and
information theory is much better suited to explaining evolution and 
the need for abstractions in our language. 
It should also be obvious that the ALP
solution to AI extends the two main tenets of logical positivism,
which are verificationism and unified science, as it gives a finite
cognitive procedure with which one can conduct all empirical work, and
allows us to develop a private language with which we can describe all
of  science and
mathematics. However, we should also mention that this strengthened
positivism does not require a strict analytic-synthetic distinction; a
spectrum of analytic-synthetic distinction as in Quine's philosophy
seems to be acceptable \cite{quine-twodogmas}. We have already seen
that according to ALP, mathematical and scientific problems have no
real distinction, therefore like Quine, ALP would allow revising even
mathematical logic itself, and we need not remind that the concept of
universal computer itself has not appeared out of thin air, but has
been invented due to the laborious mental work of scientists, as they
abstracted from the mechanics of \emph{performing mathematics}; at the
bottom these are all empirical problems \cite{chaitin-philosophical}.
On the other hand, a ``web of belief'' as in Quine, by no means
suggests non-reductionism, for that could be true only if
indeed there were phenomena that had unscathable (infinite)
complexity, such as Turing oracle machines which were not proposed as
physical machines, but only as a hypothetical concept
\cite{turing-computable}. 
Quine himself was a physicalist; 
we do not think that he would support the later vendetta against
reductionism which may be a misunderstanding of his holism. Though, it
may be argued that his obscure version of Platonism, which does not seem
much scientific to us, may be the culprit. Today's
Bayesian networks seem to be a good formalization of Quine's web of
belief, and his instrumentalism is consistent with the ALP approach of
maintaining useful programs.  Therefore, on this account, psychology
ought to be reducible to neurophysiology, as the concept of life to
molecular biology, because these are all ultimately sets of problems
that overlap in the physical world, and the relation between them
cannot hold an infinite amount of information; which would  
require an infinitely complex local environment, and that does not seem
consistent with our scientific observations.  
That is to say, discovery of bridge disciplines
is possible as exemplified by quantum chemistry and molecular biology,
and it is not different from any other kind of empirical work.
Recently, it has been perhaps better understood in the popular culture
that creationism and non-reductionism are almost synonymous (regarding
the claims of ``intelligent design'' that the flagella of bacteria are too
complex to have evolved).  Note that ALP has no qualms with the
statistical behavior of quantum systems, as it allows non-determinism.
Moreover, the particular kind of irreducibility in AIT corresponds
to weak emergentism, and most certainly contradicts with strong
emergentism which implies supernatural events. Please see also
\cite[Section 7]{dowe-survey}  for a discussion of philosophical
problems related to algorithmic complexity.

\vspace*{-6pt}
\section{Intellectual Property Towards Infinity Point}

\vspace*{-4pt}
Solomonoff has proposed the infinity point hypothesis, also known as the
singularity, as an exponentially accelerating technological progress 
caused by human-level AI's that complement the scientific community, 
to accelerate our progress ad infinitum within a finite, short time
(in practice only a finite, but significant
factor of improvement could be expected) in 1985
\cite{solomonoff-infinity} (the first paper on the subject). 
Solomonoff has proposed seven milestones
of AI development: A: modern AI phase (1956 Dartmouth conference), B:
general theory of problem solving (our interpretation: Solomonoff
Induction, Levin Search), C: self-improving AI (our interpretation:
Alpha architecture, 2002), D: AI that can understand English (our
interpretation: not realized yet), E: human-level AI, F: an AI at the
level of entire computer science (CS) community, G: an AI many times
smarter than the entire CS community.

A weak condition for the infinity point may be obtained by an economic
argument, also covered in \cite{solomonoff-infinity} briefly.  The
human brain produces 5 teraflops/watt roughly.  The current
incarnation of NVIDIA's General Purpose Graphics Programming Unit
architectures called Fermi achieves about 6 gigaflops/watt
\cite{nvidia-fermi}. Assuming 85\% improvement in power efficiency per
year (as seen in NVIDIA's projections), in 12 years, human-level
energy efficiency of computing will be achieved.  After that date, even if
mathematical AI fails due to an unforeseen problem, we will be able to
run our brain simulations faster than us, using less energy than
humans, effectively creating a bio-information based AI which meets
the basic requirement of infinity point. For this to occur, whole
brain simulation projects must be comprehensive in operation and
efficient enough \cite{sandberg-brain}.  Otherwise, human-level AI's
that we will construct should match the computational
efficiency of the human brain.  This weaker condition rests on an
economic observation: the economic incentive of cheaper intellectual
work will drive the proliferation of personal use of brain
simulations. According to NVIDIA's projections, thus, we can expect
the necessary conditions for the infinity point to materialize by
$2023$, after which point technological progress may accelerate very
rapidly. According to a recent paper by Koomey, the energy efficiency
of computing is doubling every $1.5$ years (about $60\%$ per year),
regardless of architecture, which would set the date at 2026
\cite{koomey-efficiency}.

Assume that we are progressing towards the hypothetical infinity
point. Then, the entire human civilization may be viewed as a global
intelligence working on technological problems. The practical
necessity of incremental learning suggests that when faced with more
difficult problems, better information sharing is required. If no
information sharing is present between researchers (i.e., different
search programs), then, they will lose time traversing overlapping
program subspaces. This is most clearly seen in the case of
\emph{simultaneous inventions} when an idea is said to be ``up in the
air'' and is invented by multiple, independent parties on near
dates. If intellectual property (IP) laws are too rigid and costly,
this would entail that there is minimal information sharing, and after
some point, the global efficiency of solving non-trivial technological
problems would be severely hampered. Therefore, to utilize the
infinity point effects better, knowledge sharing must be encouraged in
the society. Maximum efficiency in this fashion can be provided by
free software licenses, and a reform of the patent system. Our view is
that no single company or organization can (or should) have a monopoly
on the knowledge resources to attack problems with truly large
algorithmic complexity (monopoly is mostly illegal presently at any
rate). We tend to think that sharing science and technology is the
most efficient path towards the infinity point.  Naturally,  free
software philosophy is not acceptable to much commercial enterprise,
thus we suggest that as technology advances, the overhead of enforcing
IP laws are taken into account. If technology starts to advance much
more rapidly, the duration of the IP protection may be shortened, for
instance, as after the AI milestone F, the bureaucracy and
restrictions of IP law may be a serious bottleneck.

\vspace*{-4pt}
\section{Conclusion}

\vspace*{-4pt}
We have mentioned diverse consequences of ALP in axiomatization of AI,
philosophy, and technological society. We have also related our own
research to Solomonoff's proposals.  We interpret ALP and AIT as a
fundamentally new world-view which allows us to bridge the gap between
complex natural phenomena and positive sciences more closely than
ever. This paradigm shift has resulted in various breakthrough
applications and is likely to benefit the society in the foreseeable
future. 
\subsection*{Acknowledgements} 
We thank anonymous reviewers, David Dowe and Laurent Orseau for their valuable
comments, which substantially improved this paper.

\bibliography{agi} \bibliographystyle{splncs}

\begin{thebibliography}{10}

\bibitem{solomonoff-progress}
Solomonoff, R.J.:
\newblock Progress in incremental machine learning.
\newblock Technical Report IDSIA-16-03, IDSIA, Lugano, Switzerland (2003)

\bibitem{chaitin1975}
Chaitin, G.J.:
\newblock A theory of program size formally identical to information theory.
\newblock J. ACM \textbf{22} (1975)  329--340

\bibitem{sol56}
Solomonoff, R.J.:
\newblock An inductive inference machine.
\newblock Dartmouth Summer Research Project on Artificial Intelligence (1956) A
  privately circulated report.

\bibitem{sol57}
Solomonoff, R.J.:
\newblock An inductive inference machine.
\newblock In: IRE National Convention Record, Section on Information Theory,
  Part 2, New York, USA (1957)  56--62

\bibitem{alp1}
Solomonoff, R.J.:
\newblock A formal theory of inductive inference, part i.
\newblock Information and Control \textbf{7}(1) (1964)  1--22

\bibitem{alp2}
Solomonoff, R.J.:
\newblock A formal theory of inductive inference, part ii.
\newblock Information and Control \textbf{7}(2) (1964)  224--254

\bibitem{solomonoff-threekinds}
Solomonoff, R.J.:
\newblock Three kinds of probabilistic induction: Universal distributions and
  convergence theorems.
\newblock The Computer Journal \textbf{51}(5) (2008)  566--570 Christopher
  Stewart Wallace (1933-2004) memorial special issue.

\bibitem{levin-search}
Levin, L.A.:
\newblock Universal sequential search problems.
\newblock Problems of Information Transmission \textbf{9}(3) (1973)  265--266

\bibitem{solomonoff-theoryandapps}
Solomonoff, R.J.:
\newblock Algorithmic probability: Theory and applications.
\newblock In Dehmer, M., Emmert-Streib, F., eds.: Information Theory and
  Statistical Learning, Springer Science+Business Media, N.Y. (2009)  1--23

\bibitem{solcomplexity}
Solomonoff, R.J.:
\newblock Complexity-based induction systems: Comparisons and convergence
  theorems.
\newblock IEEE Trans. on Information Theory \textbf{IT-24}(4) (1978)  422--432

\bibitem{solomonoff-optimum}
Solomonoff, R.J.:
\newblock Optimum sequential search.
\newblock Technical report, Oxbridge Research, Cambridge, Mass., USA (1984)

\bibitem{alpstrong}
Solomonoff, R.J.:
\newblock Algorithmic Probability -- Its Discovery -- Its Properties and
  Application to Strong AI.
\newblock In: Randomness Through Computation: Some Answers, More Questions.
  World Scientific Publishing Company (2011)  149--157

\bibitem{solomonoff-incremental}
Solomonoff, R.J.:
\newblock A system for incremental learning based on algorithmic probability.
\newblock In: Proceedings of the Sixth Israeli Conference on Artificial
  Intelligence, Tel Aviv, Israel (1989)  515--527

\bibitem{rathmaner-philosophical}
Rathmanner, S., Hutter, M.:
\newblock A philosophical treatise of universal induction.
\newblock Entropy \textbf{13}(6) (2011)  1076--1136

\bibitem{davis-universalcomputer}
Davis, M.:
\newblock The Universal Computer: The Road from Leibniz to Turing.
\newblock W. W. Norton \& Company (2000)

\bibitem{turing-computable}
Turing, A.M.:
\newblock On computable numbers, with an application to the
  entscheidungsproblem.
\newblock Proceedings of the London Mathematical Society \textbf{s2-42}(1)
  (1937)  230--265

\bibitem{dowe-survey}
Dowe, D.L.:
\newblock MML, hybrid Bayesian network graphical models, statistical
  consistency, invariance and uniqueness.
\newblock In: Handbook of the Philosophy of Science - (HPS Volume 7) Philosophy
  of Statistics. Elsevier (2011)  901--982

\bibitem{WallaceBoulton:1968}
Wallace, C.S., Boulton, D.M.:
\newblock A information measure for classification.
\newblock Computer Journal \textbf{11}(2) (1968)  185--194

\bibitem{Wallace:05}
Wallace, C.S.:
\newblock Statistical and Inductive Inference by {M}inimum {M}essage {L}ength.
\newblock Springer, Berlin, Germany (2005)

\bibitem{barron2000}
Barron, A., Rissanen, J., Yu, B.
\newblock In: The minimum description length principle in coding and modeling
  (invited paper). IEEE Press, Piscataway, NJ, USA (2000)  699--716

\bibitem{vapnik-statistical}
Vapnik, V.:
\newblock Statistical Learning Theory.
\newblock John Wiley and Sons, NY (1998)

\bibitem{Dowe2007}
Dowe, D.L., Gardner, S., Oppy, G.:
\newblock Bayes not bust! why simplicity is no problem for bayesians.
\newblock The British Journal for the Philosophy of Science \textbf{58}(4)
  (2007)  709--754

\bibitem{wallace99}
Wallace, C.S., Dowe, D.L.:
\newblock Minimum message length and kolmogorov complexity.
\newblock The Computer Journal \textbf{42}(4) (1999)  270--283

\bibitem{orallo2010}
HernÃ¡ndez-Orallo, J., Dowe, D.L.:
\newblock Measuring universal intelligence: Towards an anytime intelligence
  test.
\newblock Artificial Intelligence \textbf{174}(18) (2010)  1508 -- 1539

\bibitem{DBLP:conf/agi/Ozkural11}
{\"O}zkural, E.:
\newblock Towards heuristic algorithmic memory.
\newblock In Schmidhuber, J., Th{\'o}risson, K.R., Looks, M., eds.: AGI. Volume
  6830 of Lecture Notes in Computer Science., Springer (2011)  382--387

\bibitem{oops}
Schmidhuber, J.:
\newblock Optimal ordered problem solver.
\newblock Machine Learning \textbf{54} (2004)  211--256

\bibitem{r5rs}
Richard~Kelsey, William~Clinger, J.R.:
\newblock Revised5 report on the algorithmic language scheme.
\newblock Higher-Order and Symbolic Computation \textbf{11}(1) (1998)

\bibitem{adate}
Olsson, J.R.:
\newblock Inductive functional programming using incremental program
  transformation.
\newblock Artificial Intelligence \textbf{74} (1995)  55--83

\bibitem{muggleton1994learnability}
Muggleton, S., Page, C.:
\newblock A learnability model for universal representations.
\newblock In: Proceedings of the 4th International Workshop on Inductive Logic
  Programming. Volume 237., Citeseer (1994)  139--160

\bibitem{ferri2001incremental}
Ferri-Ram{\'\i}rez, C., Hern{\'a}ndez-Orallo, J., Ramirez-Quintana, M.:
\newblock Incremental learning of functional logic programs.
\newblock FLOPS '01: Proceedings of the 5th International Symposium on
  Functional and Logic Programming (2001)  233--247

\bibitem{bowers2001knowledge}
Bowers, A., Giraud-Carrier, C., Lloyd, J., SA, E.:
\newblock A knowledge representation framework for inductive learning (2001)

\bibitem{kitzelmann2010inductive}
Kitzelmann, E.:
\newblock Inductive programming: A survey of program synthesis techniques.
\newblock Approaches and Applications of Inductive Programming (2010)  50--73

\bibitem{Looks:2007}
Looks, M.:
\newblock Scalable estimation-of-distribution program evolution.
\newblock In: Proceedings of the 9th annual conference on Genetic and
  evolutionary computation. (2007)

\bibitem{Hutter:07aixigentle}
Hutter, M.:
\newblock Universal algorithmic intelligence: A mathematical
  top$\rightarrow$down approach.
\newblock In Goertzel, B., Pennachin, C., eds.: Artificial General
  Intelligence. Cognitive Technologies.
\newblock Springer, Berlin (2007)  227--290

\bibitem{Goertzel09a}
Goertzel, B.:
\newblock Opencogprime: A cognitive synergy based architecture for artificial
  general intelligence.
\newblock In Baciu, G., Wang, Y., Yao, Y., Kinsner, W., Chan, K., Zadeh, L.A.,
  eds.: IEEE ICCI, IEEE Computer Society (2009)  60--68

\bibitem{Schmidhuber:09gm}
Schmidhuber, J.:
\newblock Ultimate cognition {\em \`{a} la} {G\"{o}del}.
\newblock Cognitive Computation \textbf{1}(2) (2009)  177--193

\bibitem{solomonoff-pastandfuture}
Solomonoff, R.J.:
\newblock Machine learning - past and future.
\newblock In: The Dartmouth Artificial Intelligence Conference. (2006)  13--15

\bibitem{Sol:97discovery}
Solomonoff, R.J.:
\newblock The discovery of algorithmic probability.
\newblock Journal of Computer and System Sciences \textbf{55}(1) (1997)  73--88

\bibitem{bridges-constructivequantum}
Bridges, D., Svozil, K.:
\newblock Constructive mathematics and quantum physics.
\newblock International Journal of Theoretical Physics \textbf{39} (2000)
  503--515

\bibitem{ozkural-compromise}
\"{O}zkural, E.:
\newblock A compromise between reductionism and non-reductionism.
\newblock In: Worldviews, Science and Us: Philosophy and Complexity. World
  Scientific Books (2007)

\bibitem{quine-twodogmas}
Quine, W.:
\newblock Two dogmas of empiricism.
\newblock The Philosophical Review \textbf{60} (1951)  20--43

\bibitem{chaitin-philosophical}
Chaitin, G.J.:
\newblock Two philosophical applications of algorithmic information theory.
\newblock In C.~S.~Calude, M. J.~Dinneen, V.V., ed.: Proceedings DMTCS'03.
  (2003)

\bibitem{solomonoff-infinity}
Solomonoff, R.J.:
\newblock The time scale of artificial intelligence: Reflections on social
  effects.
\newblock Human Systems Management \textbf{5} (1985)  149--153

\bibitem{nvidia-fermi}
Glaskowsk, P.N.:
\newblock Nvidia's fermi: The first complete gpu computing architecture (2009)

\bibitem{sandberg-brain}
Sandberg, A., Bostrom, N.:
\newblock Whole brain emulation: A roadmap.
\newblock Technical report, Future of Humanity Institute, Oxford University
  (2008)

\bibitem{koomey-efficiency}
Koomey, J.G., Berard, S., Sanchez, M., Wong, H.:
\newblock Implications of historical trends in the electrical efficiency of
  computing.
\newblock IEEE Annals of the History of Computing \textbf{33} (2011)  46--54

\end{thebibliography}

\end{document}